\documentclass[floatfix,%
 aip,
 amsmath,amssymb,
 reprint,%
]{revtex4-1}

\usepackage{graphicx}
\usepackage{dcolumn}
\usepackage{bm}
\usepackage[utf8]{inputenc}
\usepackage[T1]{fontenc}
\usepackage{mathptmx}
\usepackage{siunitx}
\usepackage{etoolbox}
\usepackage{textcomp}
\usepackage{gensymb}
\usepackage{tabularx}
\usepackage{booktabs}
\usepackage{xr}
\makeatletter
\newcommand*{\addFileDependency}[1]{
  \typeout{(#1)}
  \@addtofilelist{#1}
  \IfFileExists{#1}{}{\typeout{No file #1.}}
}
\makeatother

\newcommand*{\myexternaldocument}[1]{
    \externaldocument{#1}
    \addFileDependency{#1.tex}
    \addFileDependency{#1.aux}
}
\myexternaldocument{SI_revised}

\usepackage[normalem]{ulem}
\usepackage[dvipsnames]{xcolor}

\usepackage[dvipsnames]{xcolor} 
\newif\ifPEHighlitedChanges
\def\ifPEHighlitedChanges{\iftrue}
\ifPEHighlitedChanges
  
  \def\SPE#1{{\color{red}\sout{#1}}}

\else
  
  \def\SPE#1{\relax}

\fi
\newif\ifJCCHighlitedChanges
\def\ifJCCHighlitedChanges{\iffalse}
\ifJCCHighlitedChanges
  \def\IJCC#1{{\color{blue}#1}}
  \def\SJCC#1{{\color{blue}\sout{#1}}}

\else
  \def\IJCC#1{#1}
  \def\SJCC#1{\relax}

\fi

\begin{document}

\title{Intrinsic volume fraction can decouple from hydrodynamic size in p(NIPAm-\textit{co}-AA) microgels}

\author{Peter Edimeh}
\affiliation{
Department of Chemical and Biomolecular Engineering,
University of Houston, Houston, Texas 77204, USA
}

\author{Howard Chen}
\affiliation{
Department of Chemical and Biomolecular Engineering,
University of Houston, Houston, Texas 77204, USA
}

\author{Jacinta C. Conrad}
\email{jcconrad@uh.edu}
\affiliation{
Department of Chemical and Biomolecular Engineering,
University of Houston, Houston, Texas 77204, USA
}

\date{\today}

\begin{abstract}
Quantifying the packing fraction of microgels is key to understanding their rheology and transport properties. Often, packing fractions are extrapolated from the hydrodynamic size and intrinsic volume fraction under the assumption that particle size and polymer volume change are proportional. Here, we test this assumption for p(N-isopropylacrylamide-\textit{co}-acrylic acid) [p(NIPAm-\textit{co}-AA)] microgels of two crosslinker concentrations (1 wt\% and 7 wt\% methylenebisacrylamide) using capillary viscometry and light scattering (static and dynamic) above and below the pKa of acrylic acid (AA). At pH 8, where AA is fully ionized, the microgels do not undergo a pronounced volume phase transition because electrostatic repulsion hinders thermal collapse, and both measures of hydrodynamic volume are equal. At pH 4, however, p(NIPAm-\textit{co}-AA) microgels are weakly charged and exhibit a volume phase transition near the lower critical solution temperature of pNIPAm; the magnitude of the size change is greater for the less-crosslinked microgel. Above the volume phase transition temperature, the volume change from hydrodynamic radius decouples from that from viscosity. The ratio of the radius of gyration and hydrodynamic radius $R_g/R_h$ of p(NIPAm-\textit{co}-AA) microgels at high temperature exceeds the value theoretically predicted for homogeneous spheres and the deviation increases with the extent of crosslinking, suggesting that a non-uniform internal distribution of polymer gives rise to the anomalous decoupling at high temperatures. Our results indicate that care must be taken in extrapolating packing fractions for multi-responsive copolymer microgels, and suggest that copolymer composition and crosslink density may be used to tune bulk properties such as rheology through the internal microstructure.
 
\end{abstract}

\maketitle

\section{Introduction}

Thermoresponsive microgel particles offer tunable and stimulus-responsive properties that are useful for applications in sensors,\cite{Choe2018, Liu2018, Sigolaeva2022} coatings,\cite{Barth2018, Keskin2019, Nigro2024gels} and controlled release.\cite{Liu2010, Herman2015, Singh2022} Effective practical use of these particles requires understanding of how the properties of the particles depend on temperature. Poly($N$-isoproprylacrylamide) [pNIPAm] microgels are perhaps the best studied thermoresponsive particles because pNIPAm exhibits a lower critical solution temperature (LCST) in the range of 30 -- 35$\degree$C,\cite{Pelton1986, Yanase2020, Kujawa2001} close to human body temperature, and is an appealing candidate for delivery of drugs and theranostics. pNIPAm microgel particles undergo a volume phase transition at a temperature near pNIPAm's LCST due to temperature-dependent changes in the interactions between the polymer and solvent.\cite{Pelton2000, Stieger2004} The precise temperature of the volume phase transition temperature (VPTT) can be tuned through the density of crosslinks within the particles,\cite{Oh1998, Varga2001,FernandezNieves2011Book,OBryan2025crosslink} which tunes the softness of the particles. 

Incorporating a second responsive polymer into pNIPAm microgels is an established route for generating dual-responsive microgels.\cite{Holmqvist2012} For example, poly(acrylic acid) (PAA) and poly(methylacrylic acid) (MAA) are weak polyelectrolytes whose charge states depend on the solution pH through the ionization of the carboxylic acid groups. Thus, adding PAA or MAA (through random copolymerization\cite{Debord2003} or as an interpenetrating network (IPN)\cite{Franco2021,Nigro2021,Buratti2025ipnvsco}) imparts pH responsive behavior to thermoresponsive pNIPAm microgels. Generally, electrostatic repulsions between the charged groups in the weak polyelectrolyte increases the size of dual-responsive microgels compared to that of pNIPAm alone.\cite{KRATZ2000137,Burmistrova2011}

The spatial distribution of the monomers in p(NIPAm-AA) microgels controls the extent of the responses to temperature and pH.\cite{Nigro2021} This distribution can be controlled through synthesis (IPN versus random copolymer\cite{Yang2016ipnvsco,Buratti2025ipnvsco} or core-shell particles\cite{Jones2000coreshell}); through the microgel composition via the monomer fractions, \cite{Hyatt2015segregation,Hyatt2017}, the degree of crosslinking (i.e., the concentration of the crosslinker),\cite{Dongxia2006,Burmistrova_Crosslinker_2011, Witt2019} and the surfactant;\cite{Weihemeier2019surfactant} and on solution conditions, including temperature, pH, and salt concentration.\cite{Hyatt2015segregation,Franco2021} As a result, carefully-designed p(NIPAm-AA) microgels have been used as stimulus-responsive catalyst supports,\cite{Bhol2020catalyst} separations, \cite{Kureha2016separation} emulsifiers,\cite{Komarova2025emulsifier} and coatings.\cite{Buratti2020film}

The tunability and stimulus responsiveness of p(NIPAm-AA) microgels also makes them ideal  model systems in which to explore the state behavior and mechanical properties of soft colloids. Like hard spheres, p(NIPAm-AA) microgels can form equilibrium crystals\cite{Meng2007}, and non-equilibrium glasses.\cite{Nigro2020,Franco2021rheology}  Unlike hard spheres, p(NIPAm-AA) (as for other soft colloids\cite{Conley2019_Microgels}) can deform when highly concentrated  and thus can pack to fractions exceeding 1. Determination of packing fractions (necessary for comparison to hard spheres and to other soft colloids) in these systems is complicated by the fact that the structure of p(NIPAm-AA) microgels can change with concentration.\cite{Holmqvist2012,Franco2021} Microgels are typically characterized by a generalized volume fraction $\zeta = nV_0$, which assumes a dilute swollen particle volume as a reference volume; here $n$ is the number density and $V_0 = \frac{4}{3}\pi R_0^3$ is the volume of a swollen particle measured under dilute conditions. In most cases $n$ cannot be measured directly. Instead, the generalized volume fraction is calculated in terms of mass concentration $c$ following
\begin{equation}
\zeta = kc,
\label{eqn:k}
\end{equation}
where $k$ is the intrinsic volume fraction conversion factor\cite{Mattsson2009,Gasser_Fernandez-Nieves_JCP2014} determined from intrinsic viscosity measurements. For pNIPAm particles, measurements of $k$ as a function of temperature align with the volume calculated from the hydrodynamic radius $R_\mathrm{h}$, \cite{Senff1999} confirming that the pNIPAm microgels undergo isotropic expansion with temperature. Only a few studies, however, have examined the relationship between the intrinsic viscosity and hydrodynamic radius in p(NIPAm-AA) microgels,\cite{Hyatt2015segregation,Hyatt2017,Franco2021} where the spatial distribution of the AA and NIPAm monomers may alter the swelling behavior.

Here, we examine the thermoresponsive behavior of random copolymer p(NIPAm-\textit{co}-AA) microgel nanoparticles at two crosslink densities above and below the pKa of acrylic acid. At high pH, where the AA groups are strongly ionized, the microgel size does not depend on temperature for either crosslink density. At low pH, where the AA groups are weakly ionized, we find that two measurements of the volume expansion --- the change in hydrodynamic radius $R_\mathrm{h}$, determined with dynamic light scattering (DLS), and $k$ --- exhibit distinct temperature dependences for the random copolymers. Using static light scattering (SLS) to determine the radius of gyration $R_\mathrm{g}$, we show that the ratio $R_\mathrm{g}/R_\mathrm{h}$ for the copolymer microgels at low pH is greater than that predicted for isotropic spheres at high temperature, \IJCC{consistent with a nonuniform} internal distribution of mass as the microgels collapse. \IJCC{Both the deviation between $R_h$ and $k$ as well as the increase in $R_\mathrm{g}/R_\mathrm{h}$ at high temperatures are greater for the more strongly crosslinked copolymer microgels.} These results show how solution conditions (pH, temperature) \IJCC{and crosslinking density} can influence volume expansion in copolymer microgels, and suggest that unusual thermoresponsive behavior can be attained through the particle composition. 

\section{Materials and Methods}

\subsection{Materials}
All chemicals were purchased from Millipore-Sigma and used without further purification unless otherwise specified.

\subsection{Microgel synthesis}

\begin{figure}[h!]
\centering
  \includegraphics[width = 3.2 in]{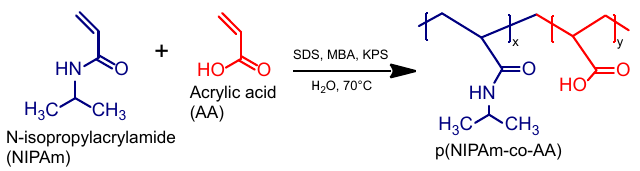}
  \caption{Synthesis scheme for the p(NIPAm-\textit{co}-AA) microgels.}

  \label{fig:Schematic1}
\end{figure}

Poly(\textit{N}-isopropylacrylamide) [pNIPAm] and poly(\textit{N}-isopropylacrylamide-\textit{co}-acrylic acid) [p(NIPAm-\textit{co}-AA)] random copolymer microgels were prepared in aqueous media through free-radical emulsion polymerization \IJCC{(Fig.\ \ref{fig:Schematic1})}, modifying procedures adapted from established pNIPAm microgel synthesis methods.\cite{Pelton2011, Mohanty_SM2016} To control the stiffness of the resulting microgels, \textit{N,N'}-methylenebisacrylamide (MBA) was introduced as a covalent crosslinker. By adding MBA at concentrations of 1 or 7 wt\% relative to the NIPAm content\footnote{The crosslinker content for the p(NIPAm-\textit{co}-AA) microgels in terms of the total polymer concentration (\%MBA$\mathrm{_{total}}$), instead of relative to NIPAm monomer (\%MBA) (Table S1), was \%MBA$\mathrm{_{total}}$ = 0.9 and 0.6 for 1 and 7 \%MBA respectively.}, we synthesized microgels with low and high crosslink densities. Potassium persulfate (KPS) served as the thermal initiator, and sodium dodecyl sulfate (SDS) was used as a surfactant to stabilize the emulsion during synthesis.

For p(NIPAm-\textit{co}-AA) synthesis, 4.540~g of NIPAm ($\ge$ 99\% purity) and 0.454~g of AA (99\% purity) were dissolved in deionized water under a nitrogen (99.9995\% purity) atmosphere. The monomer solution was filtered through a 0.22~$\mu$m syringe filter directly into a 500~mL three-neck round-bottom flask (with an opening for a nitrogen inlet, a condenser inlet, and inlet for charging in feed into the flask). Separately, 0.59~g of SDS ($\ge$ 99\% purity) was dissolved in deionized water under nitrogen and then transferred to the reaction flask. A certain amount of MBA (99\% purity) was subsequently added, with the amount of MBA depending on the target crosslink density. The reaction mixture was heated to 70~$^{\circ}$C and continuously purged with nitrogen for 1~h under constant stirring to remove dissolved oxygen. Polymerization was initiated by adding 0.178~g of KPS ($\ge$ 99\% purity). Shortly after adding the initiator, the solution turned milky white, which indicated the onset of the microgel particle formation. The reaction was allowed to proceed for 6~h at 70~$^{\circ}$C under a nitrogen atmosphere. pNIPAm microgels were synthesized following the same procedure used for p(NIPAm-\textit{co}-AA) microgel synthesis, except that no acrylic acid monomer was added to the reaction mixture.

After the reaction was completed, the microgel suspension was cooled to room temperature and purified by dialysis against ultrapure Milli-Q water (18.2~M$\Omega \cdot$cm, Millipore) \IJCC{using a 12 -- 14kDa MWCO dialysis membrane (Millipore Sigma)} to remove unreacted monomers, residual surfactant, and initiators. Dialysis was continued with daily DI water change until the conductivity of the external water \IJCC{was constant and equal} that of fresh Milli-Q water \IJCC{within measurement error} (usually 6 -- 8 days). \IJCC{We measured the pH during and after synthesis for the 7\% MBA p(NIPAM-\textit{co}-AA microgels; the pH was 3.57 $\pm$ 0.01 during synthesis and 3.52 $\pm$ 0.01 after dialysis.} The purified microgels were freeze-dried using a Labconco FreeZone lyophilizer and stored in airtight containers within desiccators at room temperature. Before each experiment, the dried microgels were redispersed in deionized water and allowed to equilibrate for 4 -- 7 days to ensure complete and homogeneous swelling.

\subsection{Elemental analysis and composition}

The composition of the p(NIPAm-\textit{co}-AA) microgels were analyzed by elemental analysis (Covalent Material Insight, CA, USA) combined with quantitative $^{1}$H NMR spectroscopy (Table {\ref{tab:elemental_analysis}). The composition of the pNIPAm and p(NIPAm-\textit{co}-AA) microgels in mol\% and determination of NIPAm and MBA content are provided in Table S2. Hereafter, we refer to the two samples using the MBA compositions in the synthesis (1 and 7 wt\%), which are close to the values determined via elemental analysis (1.57\% and 6.33\%).

\begin{table}[htbp]
    \centering
    \caption{Elemental analysis and effective compositions of the p(NIPAm \textit{co}-AA) microgels.}
    \label{tab:elemental_analysis}
    \resizebox{\columnwidth}{!}{%
    \begin{tabular}{lcccccccc}
    \toprule
    & \multicolumn{4}{c}{Elemental analysis (wt\%)}
    & \multicolumn{3}{c}{Effective composition (wt\%)} \\
    \cmidrule(lr){2-5}
    \cmidrule(lr){6-8}
    MBA
    & C
    & H
    & N
    & O
    & NIPAm
    & AA
    & MBA$\mathrm{_{total}}$ \\
    \midrule
    1\%
    & 57.64
    & 8.93
    & 10.37
    & 22.49
    & 87.95
    & 10.47
    & 1.57 \\
    
    7\%
    & 56.73
    & 8.73
    & 10.49
    & 23.83
    & 82.75
    & 10.92
    & 6.33 \\
    \bottomrule
\end{tabular}}
\end{table}

\subsection{Proton nuclear magnetic resonance spectroscopy ($^{1}$H NMR)}

$^{1}$H NMR spectra of the pNIPAm and p(NIPAm-\textit{co}-AA) microgels were obtained using a JEOL ECA-600 spectrometer operating at 600 MHz. Samples were dispersed in deuterated water (D$_2$O) and measured at room temperature. The resulting spectra were used to confirm the chemical composition (Tables \ref{tab:elemental_analysis}, S2) and molecular structure of the synthesized microgels\cite{Cooperstein2013,Buratti2025ipnvsco} (Fig.\ \ref{fig:NMR_FTIR}(a) and Table S3). 

\begin{figure}[h!]
\centering
  \includegraphics[width = 3.2 in]{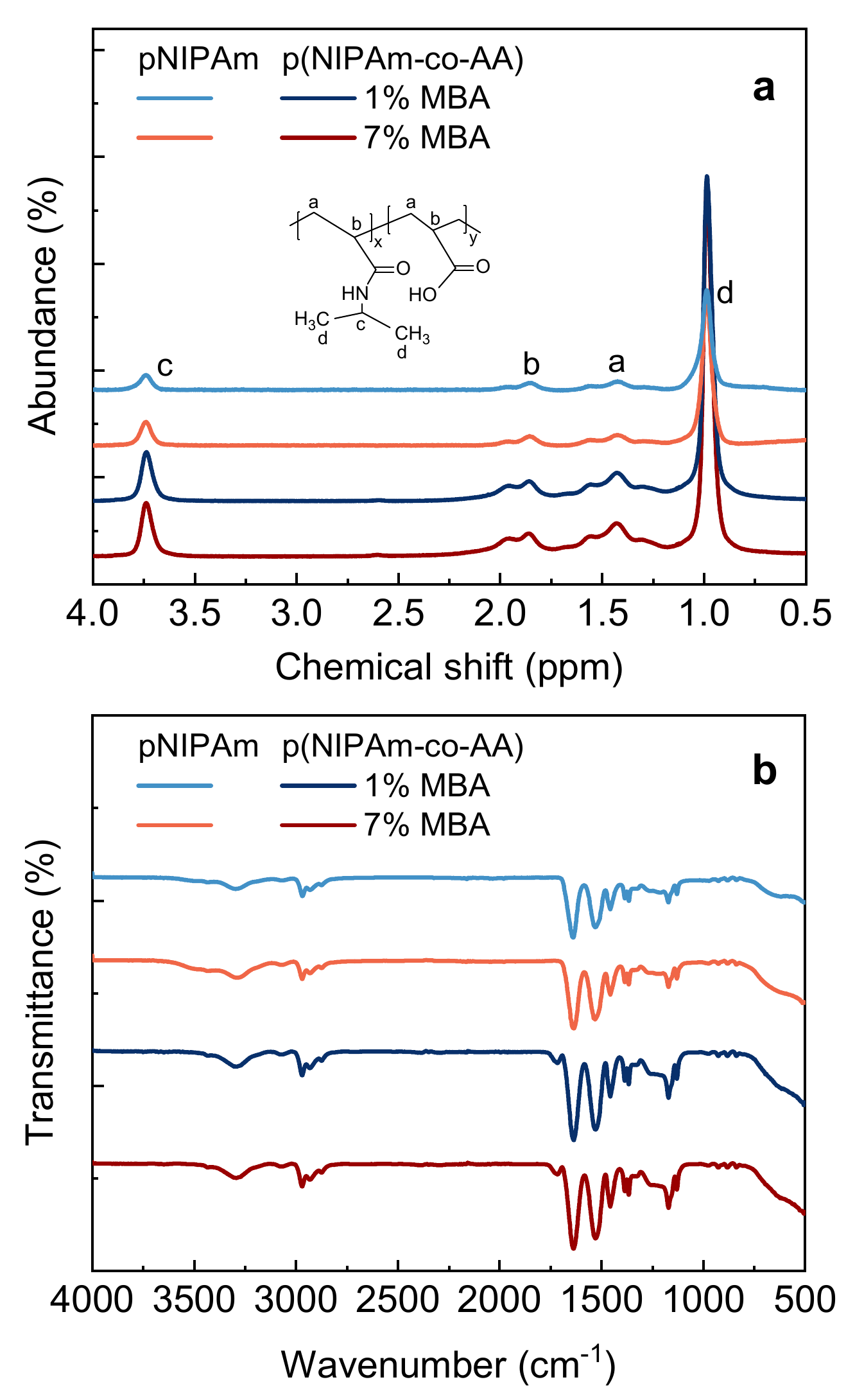}
   \caption{Chemical characterization of the microgels. (a) NMR characterization of pNIPAm and p(NIPAm-\textit{co}-AA) microgels. (b) FTIR characterization of pNIPAm and p(NIPAm-\textit{co}-AA) microgels. The NMR peaks (Table S3) are assigned following Refs.\ \citenum{Cooperstein2013,Buratti2025ipnvsco}. FTIR peak assignments are provided in Table S4.}
  \label{fig:NMR_FTIR}
\end{figure}

\subsection{Fourier transform infrared spectroscopy (FTIR)}
Infrared absorbance spectra of the pNIPAm and p(NIPAm-\textit{co}-AA) copolymer microgels were recorded using a Nicolet iS10 FTIR spectrometer fitted with an attenuated total reflectance (ATR) accessory. Spectral data were collected with OMNIC acquisition software by averaging 128 scans per sample at a spectral resolution of 8 cm$^{-1}$. FTIR analysis was performed to verify the incorporation of characteristic functional groups within the copolymer microgels, including the carboxyl groups associated with AA and the amide groups of NIPAm\cite{DYBAL200944, Kim2019} (Fig.\ \ref{fig:NMR_FTIR}(b) and Table S4).

\subsection{Static light scattering}

The radius of gyration $R_\mathrm{g}$ of the p(NIPAm-\textit{co}-AA) microgels in dilute suspensions was determined using static light scattering (SLS) using an ALV/CGS-3 compact goniometer system equipped with an ALV-7004 correlator, a He-Ne laser operating at $\lambda_0 = 632.8~\mathrm{nm}$, and a photodiode detector. For each sample, we measured the scattered intensities at angles 20 -- 150$\degree$ in $10\degree$ increments for dilute microgel suspensions with mass concentrations \IJCC{between $2.5\times10^{-3}$ and $1.5\times10^{-1}~\mathrm{g~L^{-1}}$}. The scattering of the solvent (0.22~$\mathrm{\mu m}$ filtered DI water) was measured under the same conditions and subtracted from the sample scattering, and the excess scattering intensity was normalized using toluene as the scattering standard. 

The scattering wavevector was calculated as $q = \left(4\pi n_0/\lambda_0\right) \sin{\left(\theta/2\right)}$, where $n_0$ is the refractive index of the solvent and $\theta$ is the scattering angle. Samples were equilibrated at each target temperature for 8 -- 10 min prior to SLS measurements. The angular and concentration dependence of the excess Rayleigh ratio, $R_\mathrm{vv}(q)$, was analyzed using the Zimm equation,\cite{Zimm1948, Xia_SLS2004}
\begin{equation}
\frac{Kc}{R_\mathrm{vv}(q)}
=
\frac{1}{M_\mathrm{w}}
\left(1+\frac{q^2R_\mathrm{g}^2}{3}\right)
+
2A_2c,
\label{eq:Zimm_eqn}
\end{equation}
where $K=\left(4\pi^2 n_0^2/N_\mathrm{A}\lambda_0^4 \right) \left(dn/dc\right)^2$ is an optical constant, $c$ is the microgel mass concentration, $M_\mathrm{w}$ is the weight-average molecular weight, $A_2$ is the second virial coefficient, $N_\mathrm{A}$ is Avogadro's number, and $dn/dc$ is the refractive index increment. Zimm plots were generated from $Kc/R_{vv}(q)$ as a function of both $q^2$ and $c$ followed by extrapolation to $q \rightarrow 0$ and $c \rightarrow 0$ (Figs.\ S2, S3). $R_\mathrm{g}$ was determined from the extrapolated $c \rightarrow 0$ point at low angles.

\subsection{Dynamic light scattering (DLS)}

The hydrodynamic diameter $d_\mathrm{h}$ of microgels in dilute suspensions ($\IJCC{c} \leq 4 \times 10^{-3}$ \IJCC{wt\%}) was determined by dynamic light scattering (DLS) \IJCC{using the ALV/CGS-3 compact goniometer system described in the previous section at angles 20 -- 150$\degree$ in $10\degree$ increments}. Samples were equilibrated at each target temperature for 10 -- 15 min prior to DLS measurements.

The time-averaged intensity autocorrelation function $g_2(q,\tau) = \langle I(q,t)I(q,t+\tau)\rangle/ \langle I(q,t)\rangle^2$ was measured as a function of lag time $\tau$, where $I(q,t)$ and $I(q,t+\tau)$ represent the scattered light intensities at times $t$ and $t+\tau$, respectively.\cite{RossMurphy1977} The electric field autocorrelation function, $g_1(q,\tau)$, was obtained from the Siegert relation,
$g_2(q,\tau)-1 = \beta |g_1(q,\tau)|^2$, 
where $\beta \approx 0.90$ is the instrumental coherence factor \IJCC{(Figures S1 -- S6)}. 

The decay of $g_1(q,\tau)$ is governed by the Brownian motion of the suspended microgels. The autocorrelation function was fit to a cumulant expansion,\cite{Frisken2001}
\begin{equation}
g_1(q,\tau) = \exp\left[-\Gamma(q)\tau + \frac{\mu_2\tau^2}{2}\right],
\end{equation}
where $\Gamma(q)$ is the average decay rate and $\mu_2(q)$ is the second cumulant associated with the distribution of decay rates. The translational diffusion coefficient was then obtained \IJCC{from the slope of $\Gamma$ versus $q^{2}$ (Figures S7 -- S12)} using \IJCC{$\Gamma(q)= Dq^2$} and the hydrodynamic diameter was calculated using the Stokes-Einstein equation,
$d_\mathrm{h} = k_\mathrm{B} T/3\pi \eta$, where $k_\mathrm{B}$ is the Boltzmann constant, $T$ is the absolute temperature, and $\eta$ is the viscosity of water. The polydispersity index was calculated as $\text{PDI} = \mu_2(q)/\Gamma^2(q)$. 

\subsection{Preparation of microgel suspensions for viscosity measurements}

For measurements of intrinsic viscosity, an appropriate mass of dried microgel was dispersed in deionized water to obtain an 8 wt\% stock suspension. Stock suspensions were mixed using a roller mixer for seven days to allow the microgels to completely redisperse and homogenize. Dilution series with starting concentrations ranging from 0.1 wt\% to 1.4 wt\%  were then prepared from the stock suspension, vortexed, and placed on the roller overnight. After the samples were homogenized, we gradually adjusted the pH of each suspension to the desired pH using 1M of NaOH or 0.1M HCl. \IJCC{Following certain earlier studies \cite{KRATZ2000137,Mohanty2026} we did not use a buffer so as to avoid adding more ions to the suspensions, as the conformation of p(NIPAm-\textit{co}-AA) microgels is known to be affected by salt. \cite{Schmidt2008} Instead, we monitored the pH of the suspensions over time to confirm pH stability. For samples prepared at pH$\sim$4, the pH was initially adjusted to $4.00 \pm 0.04$; over a one-week period, the pH remained between 3.95 and 4.05. For samples prepared at pH$\sim$8, the pH was initially adjusted to $8.00 \pm 0.15$; over a one-week period, the pH remained between 7.92 and 8.17.}

\subsubsection{Viscosity measurements}

The viscosity of the microgel suspensions was measured using capillary viscometry (CANNON-Ubbelohde 0B, OC, 1, 1B, and 1C viscometers) immersed in a temperature-controlled water bath. Microgel suspensions of varying concentration in vials were first equilibrated in a temperature-controlled bath for at least 30 minutes. Each sample was subsequently transferred to the viscometer, which was immersed in a second water bath maintained at the same temperature as the first bath. The sample was allowed to equilibrate inside the viscometer for at least 10 min before  measurements were performed. The efflux time was recorded for at least two (typically, three) measurements from each sample at each temperature.

\subsection{Intrinsic volume fraction determination}

We determined the intrinsic volume fraction conversion factor $k$ (eqn.\ \ref{eqn:k}) by fitting the viscosity $\eta$ using the Einstein-Batchelor expression for colloidal dispersions,\cite{Batchelor1977}
\begin{equation}
\eta = \eta_\mathrm{w} \left(1 + 2.5\zeta + 5.9\zeta^2 \right),
\label{eqn:eb}
\end{equation}
where $\eta_\mathrm{w}$ is the viscosity of water, $\zeta = kc$ is the generalized volume fraction and $c$ is the mass concentration of dried microgels.\cite{Brady1995,Mattsson2009} We determined $\eta_\mathrm{w}$ from the empirical Watson–Basu–Sengers equation\cite{Watson1980} for the dynamic viscosity of water and converted to kinematic viscosity using the temperature-dependent water density; the values obtained from this equation were in good agreement with the values that we obtained by directly fitting $\eta_\mathrm{w}$.

\section{Results and Discussion}

\subsection{Hydrodynamic diameter}

To determine how the microgels swell as a function of temperature, pH, and crosslinking density, we measured the temperature-dependent hydrodynamic diameter $d_\mathrm{h}(T)$ for 1\% and 7\% MBA p(NIPAm-\textit{co}-AA) microgels and pNIPAm microgels using DLS. The DLS correlation functions are well fit with a single-exponential cumulant model\cite{Frisken2001} across all conditions studied. Generally, the thermoresponsive size of the microgels depends on the crosslink density of the microgel network (for all microgels) and the ionization state of the acrylic acid moieties (for the random copolymer microgels).

For the pNIPAm microgels $d_\mathrm{h}$ decreases with increasing temperature $T$ (Fig.\ \ref{fig:Microgel_dH}). The size sharply decreases near 32$^\circ$C, consistent with the volume phase transition reported for pNIPAm microgels.\cite{Senff1999, FernandezNieves2011Book, Sierra-Mart_PhysRevE2011} Increasing the concentration of the MBA crosslinker from 1\% to 7\% leads to a decrease in $d_\mathrm{h}$ for the swollen microgels (at low temperature) and shifts the (apparent) temperature at which the microgels collapse to slightly higher $T$. We fit $d_\mathrm{h}$ versus $T$ to a sigmoidal function (Table S5) and determine the VPTT transition temperature to be \IJCC{31.0$\degree$C} and \IJCC{34.2$\degree$C} for the 1\% and 7\% MBA microgels, respectively. The decrease in size with increasing crosslink density reflects the increase in elastic constraints and is consistent with earlier studies of pNIPAm microgels.\cite{Oh1998, Senff2000}

\begin{figure}[h!]
\centering
  \includegraphics[width = 3.2 in]{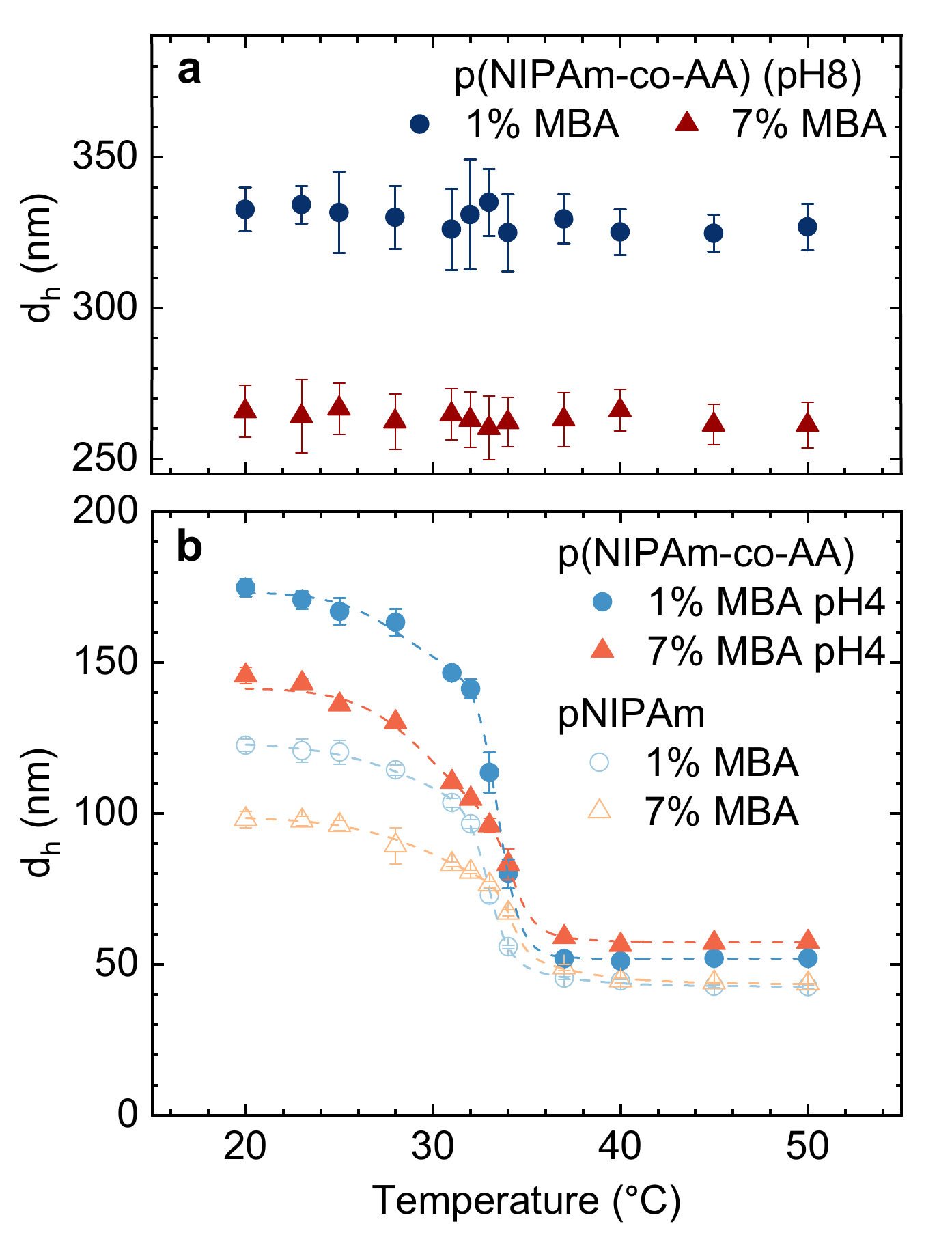}
  \caption{Hydrodynamic diameter $d_\mathrm{h}$ of p(NIPAm-\textit{co}-AA) microgels with MBA concentration of 1\% (closed blue circles) and 7\% (closed red triangles) at (a) pH 8 and (b) pH 4 as a function of temperature. In (b), $d_\mathrm{h}$ for pNIPAm microgels as a function of temperature (1\% MBA: open blue circles; 7\% MBA: open red triangles) is also presented. Dashed lines represent fits to a sigmoidal model \IJCC{and error bars indicate the standard deviation from independent measurements}.}
  \label{fig:Microgel_dH}
\end{figure}

The size of the p(NIPAm-\textit{co}-AA) microgels depends on pH as well as temperature. At pH 8, where the AA groups are nearly completely (>99\%) ionized, the hydrodynamic diameter of the p(NIPAm-\textit{co}-AA) microgels depends very weakly on the temperature  (Fig.\ \ref{fig:Microgel_dH}(a)). For example, $d_\mathrm{h}$ for the 1\% MBA copolymer microgels decreases from ca.\ 340 nm at lower temperatures ($T < 30^\circ$C) to ca.\ 330 nm for higher temperatures ($T > 33^\circ$C). A similarly subtle decrease in $d_\mathrm{h}$ is observed for the 7\% MBA microgels. The near-independence of the size on temperature is consistent with that reported at high pH for random copolymer p(NIPAm-\textit{co}-AA) microgels.\cite{Hyatt2017,Buratti2025ipnvsco} The microgels of Ref.\ \citenum{Hyatt2017} contain 24 mol\% AA and are substantially larger ($d_\mathrm{h}$ ca.\ 1.4 $\mu$m); those of Ref.\ \citenum{Buratti2025ipnvsco} contain 13 mol\% AA, similar to our system (ca.\ 16 mol\% AA), and are of very similar size ($d_\mathrm{h}$ ca.\ 350 nm). 

The lack of thermoresponsive change in size at high pH is perhaps surprising given the high pNIPAm content in our random copolymer microgels. Following Ref.\ \citenum{Hyatt2015segregation}, we estimate the probability of having at least the ten consecutive NIPAm monomers between crosslink points $\mathcal{P}(\ell_n \geq 10)$ required to observe the thermoresponsive collapse in pNIPAm chains.\cite{Ahmed2009minimum} Using the effective molar compositions, we determine $\mathcal{P}(\ell_n \geq 10)$ to be 98\% for the 1\% MBA p(NIPAm-\textit{co}-AA) microgels and 46\% for the 7\% MBA microgels (Table S6). These estimates suggest that both microgels contain sufficiently long NIPAm-rich sequences that are capable of local collapse. Indeed, such sequences are essentially guaranteed in the 1\% MBA microgels by this estimate and are more likely than not in 7\% microgels.

Instead, we suggest that the lack of variation in $R_\mathrm{h}$ with temperature is a consequence of local microphase separation in the microgel. Such a scenario is predicted by a microscopic statistical theory for partially ionized polyelectrolytes due to Borue and Erukhimovich (BE) \cite{Borue1988theory} under poor solvent conditions and confirmed by Ref.\ \citenum{Hyatt2017} for random p(NIPAm-\textit{co}-AA) microgels. In the BE theory, the state diagram is determined by two parameters: a reduced charge concentration $s$ and a reduced temperature $t$. Using the BE theory\cite{Borue1988theory} as applied to microgels,\cite{Hyatt2017} we calculated $s = 0.073$ and $0.109$ for 1 and 7\% microgels, respectively (Table S7). Because the microgels deswell at high temperatures, we assume\cite{Hyatt2017} poor solvent conditions ($t<0$). According to the BE theory, microphase separation occurs when $s<1$ and $t<1$, as found in our system. The physics of the microscopic phase separation is that the volume change driven by demixing of pNIPAm at high temperatures is opposed by the entropic penalty of confining the counterions associated with the AA monomers; to partially offset this cost, the network swells locally near the charged groups. Thus, the concentration of polymer is non-uniform inside the microgel. The insensitivity of $R_\mathrm{h}$ to temperature (Fig.\ \ref{fig:Microgel_dH}(a)) is the hallmark of this microphase separation.\cite{Hyatt2017} 

By contrast, at pH 4 the $d_\mathrm{h}$ of the p(NIPAm-\textit{co}-AA) microgels decreases with temperature. Here, below the p$K_a$ = 4.4 of acrylic acid, the AA groups are mostly (72\%) protonated (i.e., un-ionized) and the electrostatic repulsions in the copolymer are reduced. The thermoresponsive behavior of pNIPAm enabled by the high fraction of long NIPAm runs in the copolymer dominates, resulting in the pronounced decrease in $d_\mathrm{h}$ with increasing temperature that signals the volume phase transition of pNIPAm. From sigmoidal fits to $d_\mathrm{h}$ versus $T$, we determine the VPTT of the p(NIPAm-\textit{co}-AA) microgels at pH 4 to be \IJCC{33.5$\degree$C and 34.2$\degree$C} for the 1\% and 7\% MBA microgels, respectively. Thus, protonation of the AA groups at pH 4 suppresses the electrostatic effect enough to restore the thermoresponsive behavior of the pNIPAm network. As expected, the copolymer microgels with 7\% MBA are smaller than those with 1\% MBA at a given temperature due to the greater crosslink density and hence elastic constraints. 

\subsection{Radius of gyration}

To characterize the distribution of mass within the copolymer microgels, we measured the radius of gyration $R_\mathrm{g}$ at pH 4 as a function of temperature using SLS. Similar to $R_\mathrm{h}$, $R_\mathrm{g}$ decreases with increasing temperature (Fig.\ \ref{fig:Microgel_Rg}). The crosslinker concentration affects the extent of change in $R_\mathrm{g}$: at low temperature, $R_\mathrm{g}$ of the softer 1\% MBA microgels is greater than that of the 7\% MBA microgels, whereas at high temperature (above the LCST of pNIPAm) $R_\mathrm{g}$ for the deswollen 1\% MBA microgels is slightly lower than that of the 7\% microgels. This crossover suggests that the polymer mass in loosely crosslinked 1\% MBA microgels is able to reorganize as the LCST of pNIPAm is exceeded, whereas the greater number of crosslinks in the 7\% MBA microgels imposes an elastic constraint on the magnitude of the collapse in the hydrodynamic radius.\cite{Oh1998, Senff2000} 

The ratio of the radius of gyration and the hydrodynamic radius $R_\mathrm{g}/R_\mathrm{h}$ is a measure of the scattering length density inside the microgel. The refractive indices of NIPAm ($\approx 1.52$)\cite{Reufer2009} and AA ($\approx 1.42$)\cite{BAMM2013AcrylicAcid} are similar; thus, this ratio can also be interpreted as a measure of the distribution of mass inside the microgel.\cite{Hyatt2017} Specifically, $R_\mathrm{g}$ is sensitive to the radial distribution of polymer mass whereas $R_\mathrm{h}$ is the overall hydrodynamic size of the particle. At low temperatures, $R_\mathrm{g}/R_\mathrm{h}$ for both the 1\% and 7\% MBA microgels is close to the value for a homogeneous sphere, $R_\mathrm{g}/R_\mathrm{h}=\sqrt{3/5} \approx 0.778$ (Fig.\ \ref{fig:Microgel_Rg} inset). Upon increasing the temperature, \IJCC{however,} $R_\mathrm{g}/R_\mathrm{h}$ \IJCC{for both the 1\% and 7\% MBA microgels} increases. \IJCC{Above the VPTT, $R_\mathrm{g}/R_\mathrm{h}$ for the more-crosslinked 7\% MBA microgels is greater than that for the less-crosslinked 1\% MBA microgels due to its larger $R_\mathrm{g}$. Further, $R_\mathrm{g}/R_\mathrm{h}$ for the 7\% MBA microgels approaches 1,} the limiting value if the polymer mass is distributed as a thin shell at the periphery; such a picture was recently theoretically predicted.\cite{Rumyantsev2015segregation} Thus, the increase in $R_\mathrm{g}/R_\mathrm{h}$ with temperature (Fig.\ \ref{fig:Microgel_Rg}(inset)) \IJCC{suggests} that the deswelling is not spatially homogeneous for either crosslink concentration. 


Ref.\ \citenum{Hyatt2015segregation} \IJCC{also found that} $R_\mathrm{g}/R_\mathrm{h}$ \IJCC{increased} with temperature \IJCC{above the VPTT} for larger p(NIPAm-\textit{co}-AAc) microgels\IJCC{; using small-angle neutron scattering (SANS), they confirmed the core-shell mass distribution and linked the increase in $R_\mathrm{g}/R_\mathrm{h}$} to a length scale associated with internal heterogeneity. Specifically, they proposed that large amounts of AA frustrates the cooperative pNIPAm coil-to-globule transition. Because the AA groups at pH 4 are expected to be only very weakly ionized, our results are consistent with their conclusion that AA incorporation promotes inhomogeneous deswelling in p(NIPAm-\textit{co}-AA) microgels.

\begin{figure}[h!]
\centering
  \includegraphics[width = 3.2 in]{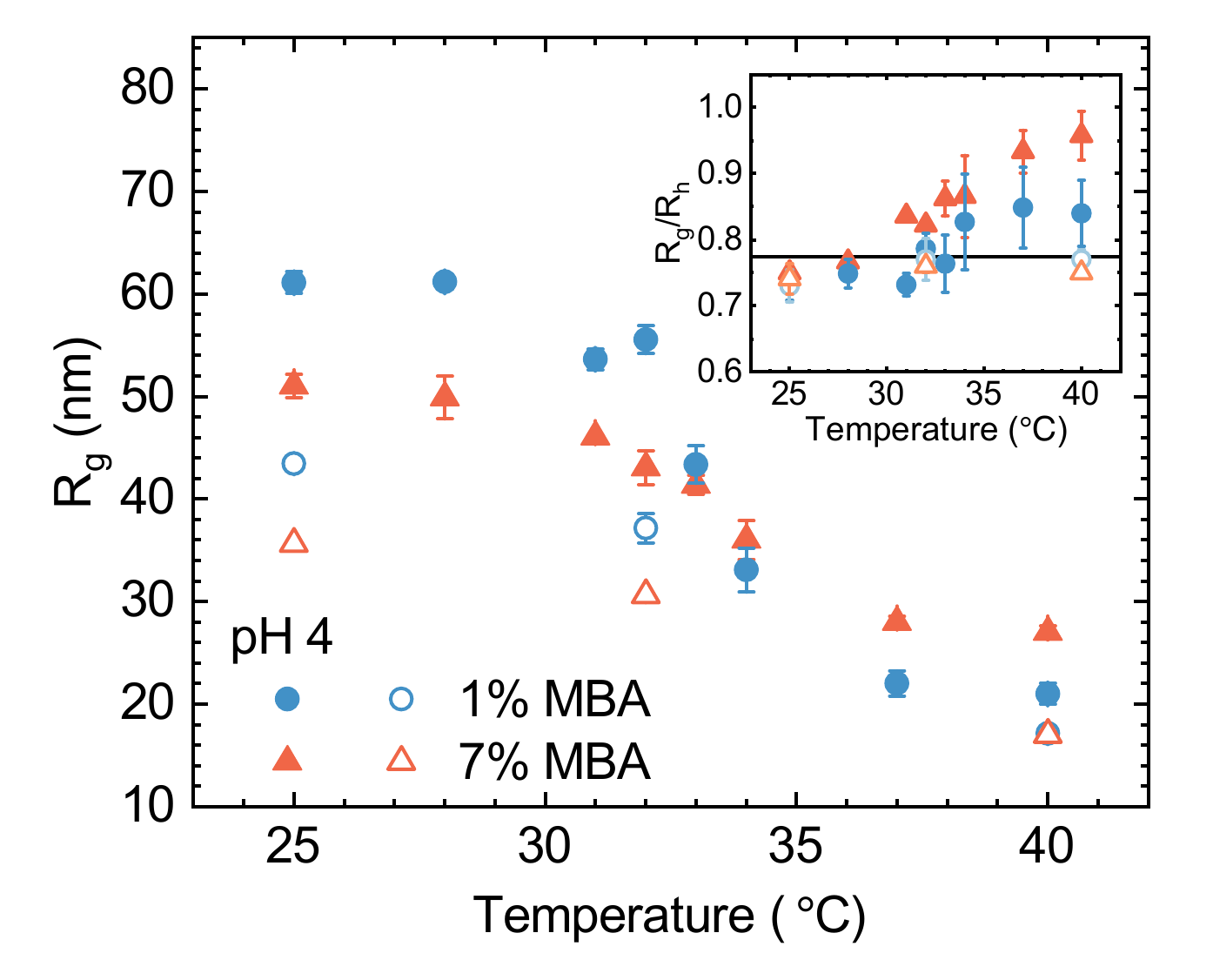}
  \caption{Radius of gyration $R_\mathrm{g}$ as a function of temperature for p(NIPAm-\textit{co}-AA) \IJCC{(closed symbols)} microgels at pH 4 and \IJCC{pNIPAm (open symbols) microgels} with MBA concentrations of 1\% (blue circles) and 7\% (orange triangles). Inset:  $R_\mathrm{g}$/$R_\mathrm{h}$ as a function of temperature for p(NIPAm-\textit{co}-AA) \IJCC{(closed symbols)} microgels at pH 4 and \IJCC{pNIPAm (open symbols) microgels} with MBA concentrations of 1\% (blue circles) and 7\% (orange triangles). Error bars indicate the uncertainty from the calculation of $R_\mathrm{g}$ and the error propagated onto $R_\mathrm{g}/R_\mathrm{h}$ \IJCC{and are smaller than the symbol size if not visible}.}
  \label{fig:Microgel_Rg}
\end{figure}

\subsection{Intrinsic viscosity}

Using intrinsic viscosity measurements, we determined the relative viscosity $\eta_\mathrm{r} = \eta / \eta_\mathrm{w}$ as a function of mass concentration $c$ for the pNIPAm and p(NIPAm-\textit{co}-AA) microgels. For brevity, we show only the results for the 1\% MBA p(NIPAm-\textit{co}-AA) microgels in the main text; those for 7\% MBA and pNIPAm are included in the SI. Generally, the relative viscosity $\eta_\mathrm{r}$ increases with $c$ as the volume occupied by the microgels increases for both pNIPAm (Fig.\ S17)  and p(NIPAm-\textit{co}-AA) microgels (Figs.\ \ref{fig:Relative_viscosity_1MBA} and S18). For all microgels, increasing the crosslinker concentration from 1\% to 7\% MBA leads to a decrease in $\eta_\mathrm{r}$. This result is consistent with a prior study that examined the effect of crosslinking density on the viscosity of pNIPAm\cite{Senff2000} microgel suspensions. 

\begin{figure}[h!]
\centering
  \includegraphics[width = 3.2 in]{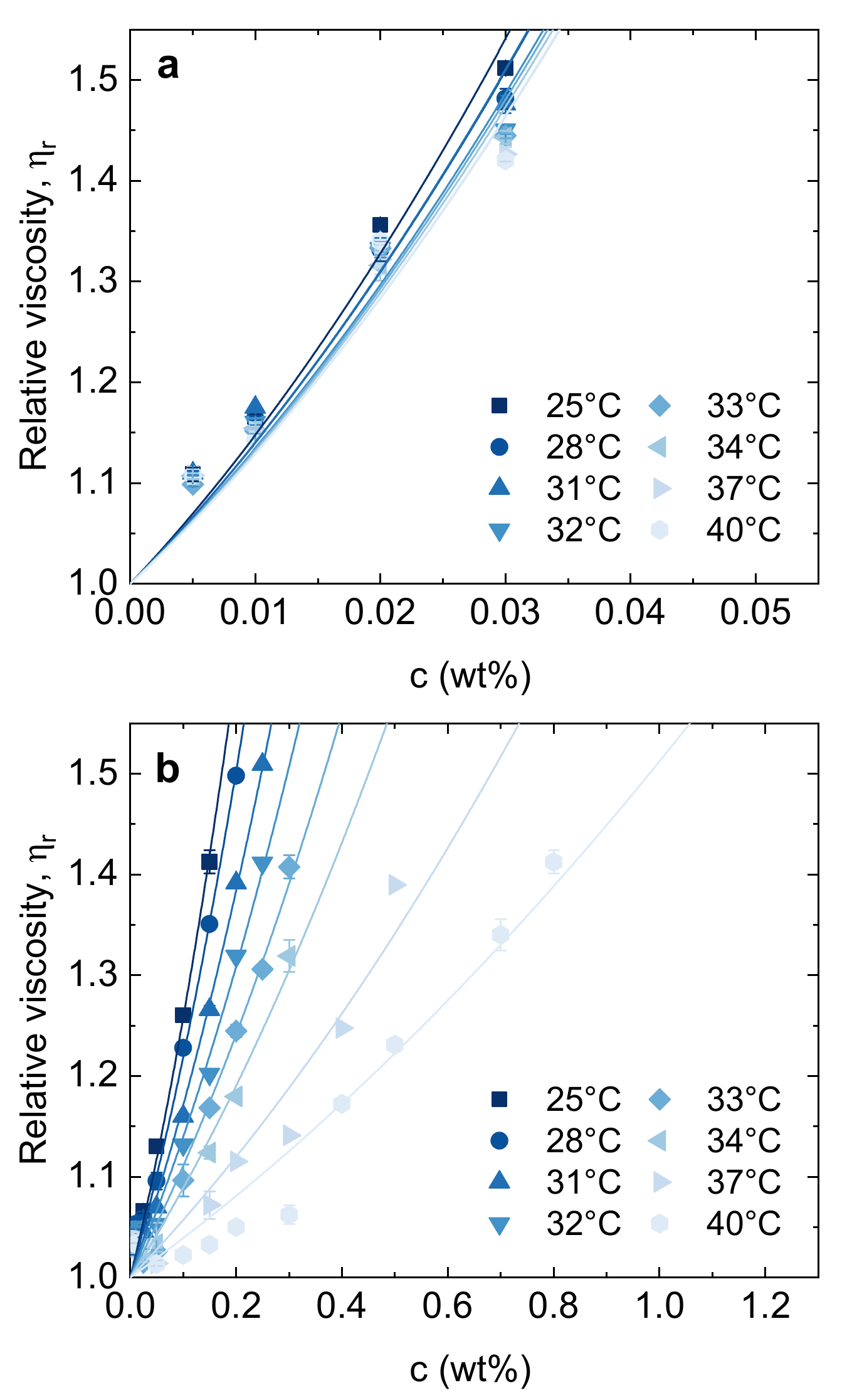}
  \caption{Relative viscosity $\eta_r$ as a function of mass concentration for 1\% MBA p(NIPAm-\textit{co}-AA) microgels at (a) pH 8 and (b) pH 4 at various temperatures, as indicated in the legend. Solid lines represent fits to the Einstein-Batchelor equation (Eqn.\ \ref{eqn:eb}) \IJCC{and error bars indicate the standard deviation from independent measurements}.}
  \label{fig:Relative_viscosity_1MBA}
\end{figure}

At pH 8, $\eta_\mathrm{r}$ for the \IJCC{1\% MBA} p(NIPAm-\textit{co}-AA) microgels is \IJCC{approximately} independent of temperature (Fig.\ \ref{fig:Relative_viscosity_1MBA}(a)). The temperature independence of the intrinsic viscosity is consistent with the lack of variation in $R_\mathrm{h}$ with temperature. Further, the $\eta_\mathrm{r}$ versus $T$ data are well fit by eqn.\ \ref{eqn:eb} at all temperatures. At pH 4, $\eta_\mathrm{r}$ for the 1\% p(NIPAm-\textit{co}-AA) microgels depends on temperature (Fig.\ \ref{fig:Relative_viscosity_1MBA}(b)). The relative viscosity is greatest at 25$^\circ$C and decreases with temperature; a similar decrease in $\eta_\mathrm{r}$ is also observed for the pNIPAm microgels (Fig.\ S17). The change in $\eta_\mathrm{r}$ with temperature reflects the thermally induced deswelling of NIPAm, which reduces the effective hydrodynamic volume of the microgels and lowers their viscosity at fixed mass concentration. At the highest temperatures, however, the $\eta_\mathrm{r}$ data for the copolymer microgels are less well fit by eqn.\ \ref{eqn:eb} at low concentrations\IJCC{, as also found in Ref.\ \citenum{Hyatt2017}}. We observed qualitatively similar trends in $\eta_\mathrm{r}$ with $T$ at high and low pH for the 7\% MBA p(NIPAm-\textit{co}-AA) microgels (Fig.\ S17). 

\subsection{Comparison of volumetric expansion}

To determine if the suspension viscosity can be described by temperature-dependent changes in particle size, we compared two normalized measures of microgel volume: the particle volume derived from DLS, $[R_h(T)/R_h(25$\degree$C)]^3$, and the viscosity-derived volume scaling factor, $k(T)/k(25$\degree$C)$. The first quantity describes changes in the effective single-particle volume that arise from isotropic swelling/deswelling of the microgels,\cite{Senff1999,Senff2000,Mohanty2017} whereas the second measure is derived from the concentration dependence of the relative viscosity and reflects many-particle contributions to the hydrodynamic and rheological responses of the suspension.\cite{Senff2000, Stieger2004Rheo, Franco2021} Equality of these two measures implies that the volume expansion is isotropic and uniform.\cite{Senff1999,Hyatt2017}

For the pNIPAm microgels, $[R_h(T)/R_h(25$\degree$C)]^3$ and $k(T)/k(25$\degree$C)$ are nearly equal across the range of temperatures examined (Fig.\ \ref{fig:Norm_hydrodynamic_volume}(a)), consistent with earlier measurements on neutral pNIPAm microgels.\cite{Senff1999} This agreement indicates that the decrease in viscosity with increasing temperature is primarily controlled by the reduction in particle volume measured by DLS. Thus, the microgels behave as expected for neutral thermoresponsive pNIPAm undergoing approximately isotropic deswelling, and the volume measured in dilute solution from DLS provides a good estimate of the volume obtained from the dilute-suspension viscosity. Increasing the crosslinker concentration from 1\% to 7\% broadens the temperature range over which the microgels collapse transition, consistent with an earlier study of pNIPAm microgels that showed that increasing crosslink density led to a more continuous volume phase transition.\cite{Senff2000}

\begin{figure}[h!]
\centering
  \includegraphics[width = 3.2 in]{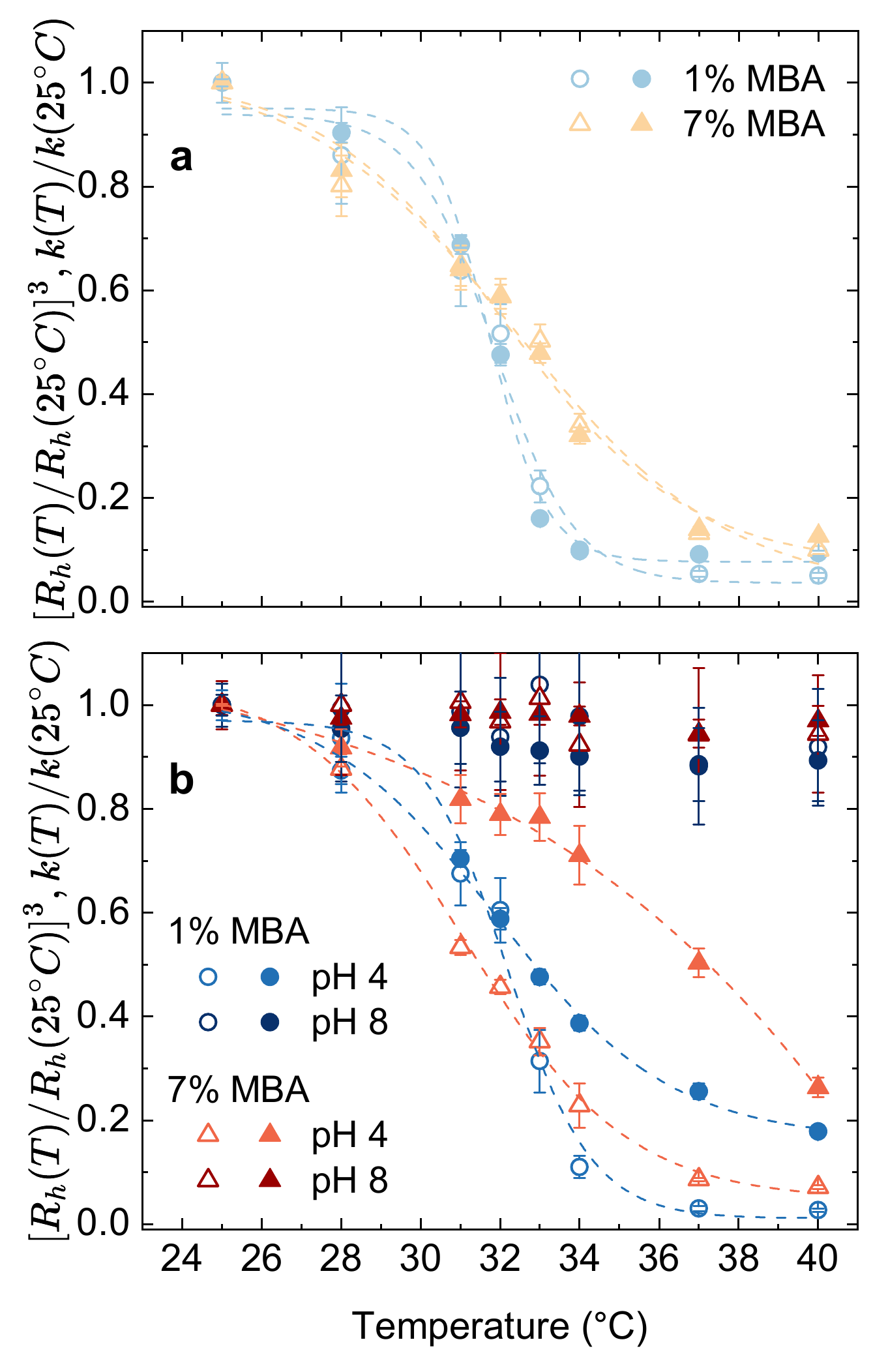}
  \caption{Normalized swelling of (a) pNIPAm and (b) p(NIPAm-\textit{co}-AA) microgels as a function of temperature with MBA concentration of 1\% (blue circles) and 7\% (orange triangles). In (b), darker symbols correspond to pH 8 and lighter symbols correspond to pH 4. Open symbols represent $[R_h(T)/R_h(25$\degree$C)]^3$ and closed symbol represent $k(T)/k(25$\degree$C)$. Dashed lines are fits to a sigmoidal model \IJCC{and error bars indicate the uncertainty from the calculations of $k$ and $R_g$ and the error propagated onto $[R_h(T)/R_h(25$\degree$C)]^3$ and $k(T)/k(25$\degree$C)$}.}
  \label{fig:Norm_hydrodynamic_volume}
\end{figure}

For p(NIPAm-\textit{co}-AA) microgels at pH 8,   $[R_h(T)/R_h(25$\degree$C)]^3$ and $k(T)/k(25$\degree$C)$ are approximately equal and at most weakly dependent on temperature (Fig.\ \ref{fig:Norm_hydrodynamic_volume}(b)). The near independence of these volume metrics from temperature contrasts with the strong temperature dependence observed for the pNIPAm microgels. The AA groups are mostly ionized at pH 8; electrostatic repulsions that appear to be strong enough to hinder p(NIPAm-\textit{co}-AA) microgels from undergoing the thermoresponsive collapse of pNIPAm near its LCST. As a result,  $[R_h(T)/R_h(25$\degree$C)]^3$ and $k(T)/k(25$\degree$C)$ do not vary significantly with temperature, and the two measures are equal within experimental error. Thus, at pH 8 the local swelling near the charged groups masks most of the temperature-induced volume change, and the effective volume probed via rheology agrees with that predicted from the particle size measured using DLS.

In striking contrast, the volumetric swelling measures for the p(NIPAm-\textit{co}-AA) microgels obtained from DLS and viscosity decouple at pH 4 for temperatures above ca.\ 30$^\circ$C (Fig.\ \ref{fig:Norm_hydrodynamic_volume}(b)). The p(NIPAm-\textit{co}-AA) particles exhibit a pronounced thermoresponsive collapse, but the decrease in $[R_\mathrm{h}(T)/R_\mathrm{h}(25$\degree$C)]^3$ (DLS) with temperature is greater than that in $k(T)/k(25$\degree$C)$ (viscosity). Thus, the hydrodynamic volume probed by viscosity does not simply scale as the cube of the hydrodynamic size measured in DLS. 
Notably, the decoupling of the two volumetric metrics occurs under conditions that lead to non-uniform mass distribution.

The two volume metrics decouple for both 1\% and 7\% MBA \IJCC{p(NIPAm-\textit{co}-AA)} microgels at pH 4. The magnitude of the volume change is \IJCC{generally} smaller for the more-crosslinked microgels in both measurements, as expected based on an earlier study on pNIPAm microgels.\cite{Senff2000} The temperature at which the two metrics decouple appears to depend slightly on the crosslinker concentration: at 31$^\circ$C the volume metrics are \IJCC{approximately equal for the 1\% MBA microgels but not for the 7\% MBA microgels}. \IJCC{The deviation between $[R_\mathrm{h}(T)/R_\mathrm{h}(25$\degree$C)]^3$ and $k(T)/k(25$\degree$C)$ is also larger for the 7\% MBA microgels than for 1\% MBA.} We speculate that \IJCC{these differences in the} temperature dependence \IJCC{and extent} of the decoupling may reflect the heterogeneity in the mass distribution inside the microgels suggested by Figure \ref{fig:Microgel_Rg}(b), which also depends on the crosslink density. 

\section{Conclusions}

We characterize the thermoresponsive properties of random copolymer p(NIPAm-\textit{co}-AA) microgels with different crosslink densities using light scattering and intrinsic viscosity measurements. At high pH, where the AA groups are nearly fully ionized, the hydrodynamic radius and relative viscosity of the microgels are nearly independent of temperature, consistent with a picture\cite{Borue1988theory,Hyatt2017} in which the polymer network locally swells near the charged groups to compensate for the entropic cost of confining the counterions as the temperature is increased. Measurements of the local volume from $R_\mathrm{h}$ and $k$ are in good agreement. At low pH the AA groups are very weakly charged, and the microgels exhibit a pronounced thermoresponsive collapse near the LCST of pNIPAm whose magnitude depends on the crosslinker concentration. Intriguingly, the two local volume measurements disagree at high temperature, which we speculate arises from the inhomogeneous distribution of mass within the microgel under these conditions indicated by values of $R_\mathrm{g}/R_\mathrm{h}$ \IJCC{that exceed $\sqrt{3/5}$ predicted for hard spheres}. \IJCC{This speculation is also supported by the predictions of BE theory, which suggest non-uniform polymer concentration; by the theoretical results of Ref.\ \citenum{Rumyantsev2015segregation}; and by comparison to the experimental results of Ref.\ \citenum{Hyatt2017}, which confirmed a core-shell conformation in p(NIPAm-\textit{co}-AA) microgels at low pH using SANS. Microgels with higher crosslink density exhibit a greater disagreement between the two local volume measurements at high temperature and greater $R_g/R_h$ values at a given temperature.}

Our results suggest that the polymer concentration within copolymer \IJCC{p(NIPAM-\textit{co}-AA)} microgels may be nonuniform\IJCC{, depending} on solution conditions \IJCC{and the extent of crosslinking}. \IJCC{An interesting open question is whether and under what conditions other charged microgels may exhibit this decoupling.} \IJCC{The} spatial inhomogeneity \IJCC{observed in this study} appears to have consequences for collective properties such as the viscosity. Thus, these results are potentially relevant for studies that use dense suspensions of microgels as models for the glass transition as well as for flow-based processing of microgel-based materials in practical applications. 

\section{Supplementary Material}

The Supplementary Material includes: \IJCC{dynamic light scattering correlation functions (Fig.\ S1 -- S6); plots of $\Gamma$ versus $q^2$ (Fig.\ S7 -- S12);} dynamic light scattering replicates (Fig.\ S13); Zimm plots from static light scattering (Figs.\ S14, S15, and S16); capillary viscometry measurements for samples not shown in main text (Figs.\ S17 and S18); determination of chemical composition (Tables S1 to S4); fitting of $d_\mathrm{h}$ to a double-sigmoidal model (Table S5)); calculation of the probability of long pNIPAM runs (Table S6); tests of the Borue-Erukhimovich theory (Table S7) ; polydispersity index (Tables S8 -- S13). 

\section{Acknowledgement}

This contribution is dedicated to our friend Murugappan Muthukumar. We thank the Welch Foundation (E-1869) and the National Science Foundation (CBET-2437846) for support. We thank Peter Vekilov for access to the ALV light scattering instrument, Alamgir Karim for access to the \IJCC{Litesizer} instrument, Ramanan Krishnamoorti for access to the FTIR equipment, Richard Willson for access to the lyophilizer, and Xudong Guan for training and access to the NMR facility. We also thank Daniel Blair, John Crocker, Ryan Poling-Skutvik, and Artem Rumyantsev for helpful discussions\IJCC{, and Amanda Marciel for assistance with the SLS analysis.}

\section{References}
\bibliography{references}

@article{Pelton1986,
title = {Preparation of aqueous latices with {N}-isopropylacrylamide},
journal = {Colloids and Surfaces},
volume = {20},
number = {3},
pages = {247-256},
year = {1986},
issn = {0166-6622},
doi = {https://doi.org/10.1016/0166-6622(86)80274-8},
author = {R.H. Pelton and P. Chibante}
}

@article{Pelton2000,
  author = {Pelton, Robert},
  title = {Temperature-Sensitive Aqueous Microgels},
  journal = {Adv. Colloid Interface Sci.},
  year = {2000},
  volume = {85},
  pages = {1-33},
  doi = {10.1016/S0001-8686(99)00023-8}
}

@inbook{Pelton2011,
author = {Pelton, Robert and Hoare, Todd},
publisher = {John Wiley and Sons, Ltd},
isbn = {9783527632992},
title = {Microgels and Their Synthesis: An Introduction},
booktitle = {Microgel Suspensions},
chapter = {1},
pages = {1-32},
doi = {https://doi.org/10.1002/9783527632992.ch1},
year = {2011},
}

@article{RossMurphy1977,
  title={Dynamic Light Scattering. B. J. Berne and R. Pecora, John Wiley, New York, 1976, pp. 376.},
  author={Ross-Murphy, Simon B.},
  journal={British Polymer Journal},
  year={1977},
  volume={9},
  pages={177-177},
  url={https://doi.org/10.1002/pi.4980090216}
}

@article{Frisken2001,
author = {Barbara J. Frisken},
journal = {Appl. Opt.},
number = {24},
pages = {4087-4091},
publisher = {Optica Publishing Group},
title = {Revisiting the method of cumulants for the analysis of dynamic light-scattering data},
volume = {40},
month = {Aug},
year = {2001},
url = {https://opg.optica.org/ao/abstract.cfm?URI=ao-40-24-4087},
doi = {10.1364/AO.40.004087},
}

@article{Mattsson2009,
  author = {Mattsson, J. and Wyss, Hans M. and Fernandez-Nieves, Alberto and Miyazaki, Kunimasa and Hu, Zhibing and Reichman, David R. and Weitz, David A.},
  title = {Soft Colloids Make Strong Glasses},
  journal = {Nature},
  year = {2009},
  volume = {462},
  pages = {83-86},
  doi = {10.1038/nature08457}
}

@article{Sierra-Mart_PhysRevE2011,
  title = {Bulk modulus of poly({$N$}-isopropylacrylamide) microgels through the swelling transition},
  author = {Sierra-Mart{\'i}n, B. and Laporte, Y. and South, A. B. and Lyon, L. A. and Fern\'andez-Nieves, A.},
  journal = {Phys. Rev. E},
  volume = {84},
  issue = {1},
  pages = {011406},
  numpages = {4},
  year = {2011},
  month = {Jul},
  publisher = {American Physical Society},
  doi = {10.1103/PhysRevE.84.011406},
  url = {https://link.aps.org/doi/10.1103/PhysRevE.84.011406}
}

@article{Batchelor1977,
    author = {{Batchelor}, G. K.},
    title = "{The effect of Brownian motion on the bulk stress in a suspension of spherical particles}",
    journal = {J. Fluid Mech.},
    year = 1977,
    month = nov,
    volume = {83},
    pages = {97-117},
    doi = {10.1017/S0022112077001062},
}

@article{Mohanty2026,
title = {Electrostatic interactions reshape the internal architecture of ionic microgels},
journal = {J. Colloid Interface Sci.},
volume = {708},
pages = {139704},
year = {2026},
issn = {0021-9797},
doi = {https://doi.org/10.1016/j.jcis.2025.139704},
url = {https://www.sciencedirect.com/science/article/pii/S0021979725030966},
author = {Priti S. Mohanty and Chi Zhang and Elisa Ballin and Francesco Brasili and Giovanni {Del Monte} and Emanuela Zaccarelli and Frank Scheffold},

}

@Article{Mohanty_SM2016,
author ="Mohanty, P. S. and Nöjd, S. and Bergman, M. J. and Nägele, G. and Arrese-Igor, S. and Alegria, A. and Roa, R. and Schurtenberger, P. and Dhont, J. K. G.",
title  ="Dielectric spectroscopy of ionic microgel suspensions",
journal  ="Soft Matter",
year  ="2016",
volume  ="12",
issue  ="48",
pages  ="9705-9727",
publisher  ="The Royal Society of Chemistry",
doi  ="10.1039/C6SM01683A",
}

@article{Hyatt2017,
  title = {Charge segregation in weakly ionized microgels},
  author = {Hyatt, John S. and Douglas, Alison M. and Stanley, Chris and Do, Changwoo and Barker, Thomas H. and Fern\'andez-Nieves, Alberto},
  journal = {Phys. Rev. E},
  volume = {95},
  issue = {1},
  pages = {012608},
  numpages = {9},
  year = {2017},
  month = {Jan},
  publisher = {American Physical Society},
  doi = {10.1103/PhysRevE.95.012608},
  url = {https://link.aps.org/doi/10.1103/PhysRevE.95.012608}
}

@article{Gasser_Fernandez-Nieves_JCP2014,
    author = {Gasser, U. and Hyatt, J. S. and Lietor-Santos, J.-J. and Herman, E. S. and Lyon, L. A. and Fernandez-Nieves, A.},
    title = {Form factor of {pNIPAM} microgels in overpacked states},
    journal = {J. Chem. Phys.},
    volume = {141},
    number = {3},
    pages = {034901},
    year = {2014},
    month = {07},
    issn = {0021-9606},
    doi = {10.1063/1.4885444},
    url = {https://doi.org/10.1063/1.4885444},
   
}

@article{Stieger2004,
author = {Stieger, Markus and Richtering, Walter and Pedersen, Jan Skov and Lindner, Peter},
title = {Small-angle neutron scattering study of structural changes in temperature sensitive microgel colloids},
journal = {J. Chem. Phys.},
volume = {120},
number = {13},
pages = {6197--6206},
year = {2004},
doi = {10.1063/1.1665752}
}

@article{Franco2021,
author = {Franco, S. and Buratti, E. and Ruzicka, B. and Nigro, V. and Zoratto, N. and Matricardi, P. and Zaccarelli, E. and Angelini, R.},
title = {Volume fraction determination of microgel composed of interpenetrating polymer networks of {PNIPAM} and polyacrylic acid},
journal = {J. Phys. Condens. Matter},
volume = {33},
number = {17},
pages = {174004},
year = {2021},
doi = {10.1088/1361-648X/abe1ec}
}

@article{Senff2000,
author = {Senff, H. and Richtering, W.},
title = {Influence of cross-link density on rheological properties of temperature-sensitive microgel suspensions},
journal = {Colloid Polym. Sci.},
volume = {278},
pages = {830--840},
year = {2000},
doi = {10.1007/s003960000329}
}

@article{Holmqvist2012,
  title = {Structure and Dynamics of Loosely Cross-Linked Ionic Microgel Dispersions in the Fluid Regime},
  author = {Holmqvist, P. and Mohanty, P. S. and N\"agele, G. and Schurtenberger, P. and Heinen, M.},
  journal = {Phys. Rev. Lett.},
  volume = {109},
  issue = {4},
  pages = {048302},
  numpages = {5},
  year = {2012},
  month = {Jul},
  publisher = {American Physical Society},
  doi = {10.1103/PhysRevLett.109.048302},
}

@article{Mohanty2017,
author = {Mohanty, Priti S. and Nöjd, Sofi and van Gruijthuijsen, Kitty and Crassous, Jérôme J. and Obiols-Rabasa, Marc and Schweins, Ralf and Stradner, Anna and Schurtenberger, Peter},
title = {Interpenetration of polymeric microgels at ultrahigh densities},
journal = {Sci. Rep.},
volume = {7},
pages = {1487},
year = {2017},
doi = {10.1038/s41598-017-01471-3}
}

@article{Nigro2020,
author = {Nigro, Valentina and Ruzicka, Barbara and Ruta, Beatrice and Zontone, Federico and Bertoldo, Monica and Buratti, Elena and Angelini, Roberta},
title = {Relaxation Dynamics, Softness, and Fragility of Microgels with Interpenetrated Polymer Networks},
journal = {Macromolecules},
volume = {53},
number = {5},
pages = {1596-1603},
year = {2020},
doi = {10.1021/acs.macromol.9b01560},
}

@book{FernandezNieves2011Book,
  editor = {Fernandez-Nieves, Alberto and Wyss, Hans M. and Mattsson, Johan and Weitz, David A.},
  title = {Microgel Suspensions: Fundamentals and Applications},
  publisher = {Wiley-VCH},
  year = {2011}
}

@article{Senff1999,
    author = {Senff, H. and Richtering, W.},
    title = {Temperature sensitive microgel suspensions: Colloidal phase behavior and rheology of soft spheres},
    journal = {J. Chem. Phys.},
    volume = {111},
    number = {4},
    pages = {1705-1711},
    year = {1999},
    month = {07},
    issn = {0021-9606},
    doi = {10.1063/1.479430},
    url = {https://doi.org/10.1063/1.479430},

}

@article{Nigro2021,
  author = {Nigro, Valentina and Angelini, Roberta and Bertoldo, Monica and Buratti, Elena and Franco, Silvia and Ruzicka, Barbara},
  title = {Chemical-Physical Behaviour of Microgels Made of Interpenetrating Polymer Networks of {PNIPAM} and Poly(acrylic Acid)},
  journal = {Polymers},
  volume = {13},
  number = {9},
  pages = {1353},
  year = {2021},
  doi = {10.3390/polym13091353},
  url = {https://www.mdpi.com/2073-4360/13/9/1353},
  issn = {2073-4360},
}

@article{Buratti2025ipnvsco,
title = {Copolymer vs interpenetrated polymer network microgels: The case of poly($N$-isopropylacrylamide) and poly(acrylic acid)},
journal = {J. Mol. Liq.},
volume = {427},
pages = {127328},
year = {2025},
issn = {0167-7322},
doi = {https://doi.org/10.1016/j.molliq.2025.127328},
url = {https://www.sciencedirect.com/science/article/pii/S0167732225004957},
author = {Elena Buratti and Silvia Franco and Giulia {Di Gregorio} and Francesca Ripanti and Valentina Nigro and Monica Bertoldo and Roberta Angelini and Paolo Postorino and Barbara Ruzicka}
}

@article{Franco2021rheology,
  author = {Franco, Silvia and Buratti, Elena and Nigro, Valentina and Zaccarelli, Emanuela and Ruzicka, Barbara and Angelini, Roberta},
  title = {Glass and Jamming Rheology in Soft Particles Made of {PNIPAM} and Polyacrylic Acid},
  journal = {Int. J. Mol. Sci.},
  volume = {22},
  number = {8},
  pages = {4032},
  year = {2021},
  doi = {10.3390/ijms22084032},
  url = {https://www.mdpi.com/1422-0067/22/8/4032},
  issn = {1422-0067},
}

@article{OBryan2025crosslink,
author = {O’Bryan, Christopher S. and Rose, Katie A. and Ford, Jamie and Ghoto, Mujtaba Rafique and Lee, Daeyeon and Composto, Russell J.},
title = {Cross-Linking Density Controls the Interstitial Pore Size and Elasticity in Polyacrylamide Microgel Packings},
journal = {Macromolecules},
volume = {58},
number = {8},
pages = {3937-3948},
year = {2025},
doi = {10.1021/acs.macromol.4c02784}
}

@article{DYBAL200944,
title = {The role of water in structural changes of poly(N-isopropylacrylamide) and poly(N-isopropylmethacrylamide) studied by FTIR, Raman spectroscopy and quantum chemical calculations},
journal = {Vib. Spectrosc.},
volume = {51},
number = {1},
pages = {44-51},
year = {2009},
issn = {0924-2031},
doi = {https://doi.org/10.1016/j.vibspec.2008.10.003},
url = {https://www.sciencedirect.com/science/article/pii/S0924203108001720},
author = {Jiří Dybal and Miroslava Trchová and Pavel Schmidt}
}

@article{Kim2019,
author  = {Kim, Yun Kyoung and Kim, Eun-Joong and Lim, Jae Hyun and Cho, Heui Kyoung and Hong, Woo Jin and Jeon, Hyang Hwa and Chung, Bong Geun},
title   = {Dual Stimuli-Triggered Nanogels in Response to Temperature and pH Changes for Controlled Drug Release},
journal = {Nanoscale Res. Lett.},
year    = {2019},
volume  = {14},
number  = {1},
pages   = {77},
doi     = {10.1186/s11671-019-2909-y},
url     = {https://doi.org/10.1186/s11671-019-2909-y}
}

@article{Liu2018,
author = {Liu, Li and Bakhshi, Hadi and Jiang, Shaohua and Schmalz, Holger and Agarwal, Seema},
title = {Composite Polymeric Membranes with Directionally Embedded Fibers for Controlled Dual Actuation},
journal = {Macromol. Rapid Commun.},
volume = {39},
number = {10},
pages = {1800082},
doi = {10.1002/marc.201800082},
url = {https://onlinelibrary.wiley.com/doi/abs/10.1002/marc.201800082},
year = {2018}
}

@article{Barth2018,
title = {Monolayer microgel composite membranes with tunable permeability},
journal = {J. Membr. Sci.},
volume = {555},
pages = {473-482},
year = {2018},
issn = {0376-7388},
doi = {https://doi.org/10.1016/j.memsci.2018.03.037},
url = {https://www.sciencedirect.com/science/article/pii/S0376738817333677},
author = {M. Barth and M. Wiese and W. Ogieglo and D. Go and A.J.C. Kuehne and M. Wessling},
}

@article{Herman2015,
author = {Herman, E. S. and Lyon, L. A.},
year = {2015},
month = {05},
pages = {1535-1544},
title = {Polyelectrolyte exchange and diffusion in microgel multilayer thin films},
volume = {293},
number = {5},
doi = {https://doi.org/10.1007/s00396-015-3547-2},
journal = {Colloid Polym. Sci.},
}

@article{Liu2010,
title = {Drug release kinetics from monolayer films of glucose-sensitive microgel},
journal = {Polymer},
volume = {51},
number = {12},
pages = {2668-2675},
year = {2010},
issn = {0032-3861},
doi = {https://doi.org/10.1016/j.polymer.2010.04.011},
author = {Pengxiao Liu and Qiaofang Luo and Ying Guan and Yongjun Zhang},

}

@article{Sigolaeva2022,
author = {Sigolaeva, Larisa V. and Pergushov, Dmitry V. and Gladyr, Snezhana Yu. and Kurochkin, Ilya N. and Richtering, Walter},
title = {Microgels in Tandem with Enzymes: Tuning Adsorption of a {pH}- and Thermoresponsive Microgel for Improved Design of Enzymatic Biosensors},
journal = {Adv. Mater. Interfaces},
volume = {9},
number = {17},
pages = {2200310},
doi = {https://doi.org/10.1002/admi.202200310},
url = {https://advanced.onlinelibrary.wiley.com/doi/abs/10.1002/admi.202200310},
year = {2022}
}

@article{Nigro2024gels,
  author = {Nigro, Valentina and Angelini, Roberta and Buratti, Elena and Colantonio, Claudia and D’Amato, Rosaria and Dinelli, Franco and Franco, Silvia and Limosani, Francesca and Montereali, Rosa Maria and Nichelatti, Enrico and Piccinini, Massimo and Vincenti, Maria Aurora and Ruzicka, Barbara},
  title = {Influence of a Solid Surface on {PNIPAM} Microgel Films},
  journal = {Gels},
  volume = {10},
  number = {7},
  pages = {473},
  year = {2024},
  doi = {10.3390/gels10070473},
  issn = {2310-2861}
}

@article{Keskin2019,
author = {Keskin, Damla and Mergel, Olga and van der Mei, Henny C. and Busscher, Henk J. and van Rijn, Patrick},
title = {Inhibiting Bacterial Adhesion by Mechanically Modulated Microgel Coatings},
journal = {Biomacromolecules},
volume = {20},
number = {1},
pages = {243-253},
year = {2019},
doi = {10.1021/acs.biomac.8b01378},
}

@article{Singh2022,
author = {Singh, Nirbhai and Aery, Shikha and Juneja, Smayira and Kumari, Laxmi and Lone, Mohd Sajid and Dar, Aijaz Ahmad and Pawar, Sandip V. and Mehta, Surinder K. and Dan, Abhijit},
title = {Chitosan Hydrogels with Embedded Thermo- and {pH}-Responsive Microgels as a Potential Carrier for Controlled Release of Drugs},
journal = {ACS Appl. Bio. Mater.},
volume = {5},
number = {7},
pages = {3487-3499},
year = {2022},
doi = {10.1021/acsabm.2c00401},
}

@article{Choe2018,
  author = {Choe, Ayoung and Yeom, Jeonghee and Shanker, Ravi and Kim, Minsoo P. and Kang, Saewon and Ko, Hyunhyub},
  title = {Stretchable and Wearable Colorimetric Patches Based on Thermoresponsive Plasmonic Microgels Embedded in a Hydrogel Film},
  journal = {NPG Asia Mater.},
  year = {2018},
  volume = {10},
  number = {10},
  pages = {912--922},
  doi = {10.1038/s41427-018-0086-6}
}

@article{Kujawa2001,
author = {Kujawa, Piotr and Winnik, Fran{\c{c}}oise M.},
title = {Volumetric Studies of Aqueous Polymer Solutions Using Pressure Perturbation Calorimetry: A New Look at the Temperature-Induced Phase Transition of Poly($N$-isopropylacrylamide) in Water and {D2O}},
journal = {Macromolecules},
volume = {34},
number = {12},
pages = {4130-4135},
year = {2001},
doi = {10.1021/ma002082h},
}

@article{Yanase2020,
author = {Keiichi Yanase and Richard Buchner and Takaaki Sato},
title = {Microscopic insights into the phase transition of poly($N$-isopropylacrylamide) in aqueous media: Effects of molecular weight and polymer concentration},
journal = {J. Mol. Liq.},
volume = {302},
pages = {112025},
year = {2020},
issn = {0167-7322},
doi = {https://doi.org/10.1016/j.molliq.2019.112025},
url = {https://www.sciencedirect.com/science/article/pii/S0167732219324262},
}

@article{Oh1998,
author = {Oh, Kwang Suk and Oh, Jeong Seok and Choi, Hyun Seok and Bae, Young Chan},
title = {Effect of Cross-Linking Density on Swelling Behavior of {NIPA} Gel Particles},
journal = {Macromolecules},
volume = {31},
number = {21},
pages = {7328-7335},
year = {1998},
doi = {10.1021/ma971554y},
}

@article{Varga2001,
author = {Varga, Imre and Gilányi, Tibor and M{\'e}száros, Róbert and Filipcsei, Genoveva and Zrínyi, Miklós},
title = {Effect of Cross-Link Density on the Internal Structure of Poly($N$-isopropylacrylamide) Microgels},
journal = {J. Phys. Chem. B},
volume = {105},
number = {38},
pages = {9071-9076},
year = {2001},
doi = {10.1021/jp004600w},
}

@article{Watson1980,
    author = {Watson, J. T. R. and Basu, R. S. and Sengers, J. V.},
    title = {An Improved Representative Equation for the Dynamic Viscosity of Water Substance},
    journal = {J. Phys. Chem. Ref. Data},
    volume = {9},
    number = {4},
    pages = {1255-1290},
    year = {1980},
    month = {10},
    issn = {0047-2689},
    doi = {10.1063/1.555631},
    url = {https://doi.org/10.1063/1.555631},
}

@article{Hyatt2015segregation,
  title = {Segregation of mass at the periphery of $N$-isopropylacrylamide-co-acrylic-acid microgels at high temperatures},
  author = {Hyatt, John S. and Do, Changwoo and Hu, Xiaobo and Choi, Hong Sung and Kim, Jin Woong and Lyon, L. Andrew and Fernandez-Nieves, Alberto},
  journal = {Phys. Rev. E},
  volume = {92},
  issue = {3},
  pages = {030302(R)},
  numpages = {5},
  year = {2015},
  month = {Sep},
  publisher = {American Physical Society},
  doi = {10.1103/PhysRevE.92.030302},
  url = {https://link.aps.org/doi/10.1103/PhysRevE.92.030302}
}

@article{Weihemeier2019surfactant,
    author = {Wiehemeier, Lars and Brändel, Timo and Hannappel, Yvonne and Kottke, Tilman and Hellweg, Thomas},
    title = {Synthesis of smart dual-responsive microgels: correlation between applied surfactants and obtained particle morphology},
    journal = {Soft Matter},
    volume = {15},
    number = {28},
    pages = {5673-5684},
    year = {2019},
    month = {07},
    issn = {1744-683X},
    doi = {10.1039/c9sm00690g}
}

@article{Jones2000coreshell,
author = {Jones, Clinton D. and Lyon, L. Andrew},
title = {Synthesis and Characterization of Multiresponsive Core-Shell Microgels},
journal = {Macromolecules},
volume = {33},
number = {22},
pages = {8301-8306},
year = {2000},
doi = {10.1021/ma001398m}
}

@article{Cooperstein2013,
author = {Cooperstein, Marta and Canavan, Heather},
year = {2013},
month = {12},
pages = {19},
title = {Assessment of cytotoxicity of ($N$-isopropyl acrylamide) and Poly($N$-isopropyl acrylamide)-coated surfaces},
volume = {8},
journal = {Biointerphases},
doi = {10.1186/1559-4106-8-19}
}

@article{Brady1995,
    author = {Brady, John F. and Vicic, Michael},
    title = {Normal stresses in colloidal dispersions},
    journal = {J. Rheol.},
    volume = {39},
    number = {3},
    pages = {545-566},
    year = {1995},
    month = {05},
    issn = {0148-6055},
    doi = {10.1122/1.550712},
    url = {https://doi.org/10.1122/1.550712},
    
}

@article{Zimm1948,
    author = {Zimm, Bruno H.},
    title = {Apparatus and Methods for Measurement and Interpretation of the Angular Variation of Light Scattering; Preliminary Results on Polystyrene Solutions},
    journal = {J. Chem. Phys.},
    volume = {16},
    number = {12},
    pages = {1099-1116},
    year = {1948},
    month = {12},
    issn = {0021-9606},
    doi = {10.1063/1.1746740},
    url = {https://doi.org/10.1063/1.1746740},

}

@article{Xia_SLS2004,
author = {Xia, Xiaohu and Hu, Zhibing},
title = {Synthesis and Light Scattering Study of Microgels with Interpenetrating Polymer Networks},
journal = {Langmuir},
volume = {20},
number = {6},
pages = {2094-2098},
year = {2004},
doi = {https://doi.org/10.1021/la0354483},
}

@article{Debord2003,
author = {Debord, Justin D. and Lyon, L. Andrew},
title = {Synthesis and Characterization of pH-Responsive Copolymer Microgels with Tunable Volume Phase Transition Temperatures},
journal = {Langmuir},
volume = {19},
number = {18},
pages = {7662-7664},
year = {2003},
doi = {10.1021/la0342924}
}

@article{Meng2007,
author = {Meng, Zhiyong and Cho, Jae Kyu and Debord, Stella and Breedveld, Victor and Lyon, L. Andrew},
title = {Crystallization Behavior of Soft, Attractive Microgels},
journal = {J. Phys. Chem. B},
volume = {111},
number = {25},
pages = {6992-6997},
year = {2007},
doi = {10.1021/jp073122n},

URL = {  https://doi.org/10.1021/jp073122n},

}

@article{Conley2019_Microgels,
  author  = {Conley, Gaurasundar M. and Zhang, Chi and Aebischer, Philippe and Harden, James L. and Scheffold, Frank},
  title   = {Relationship between rheology and structure of interpenetrating, deforming and compressing microgels},
  journal = {Nat. Commun.},
  year    = {2019},
  volume  = {10},
  number  = {1},
  pages   = {2436},
  doi     = {10.1038/s41467-019-10181-5},
}

@article{KRATZ2000137,
title = {Influence of charge density on the swelling of colloidal poly($N$-isopropylacrylamide-co-acrylic acid) microgels},
journal = {Colloids Surf. A},
volume = {170},
number = {2},
pages = {137-149},
year = {2000},
issn = {0927-7757},
doi = {https://doi.org/10.1016/S0927-7757(00)00490-8},
url = {https://www.sciencedirect.com/science/article/pii/S0927775700004908},
author = {Karl Kratz and Thomas Hellweg and Wolfgang Eimer}
}

@Article{Burmistrova2011,
AUTHOR = {Burmistrova, Anna and Richter, Marcel and Eisele, Michael and Üzüm, Cagri and Von Klitzing, Regine},
TITLE = {The Effect of Co-Monomer Content on the Swelling/Shrinking and Mechanical Behaviour of Individually Adsorbed {PNIPAM} Microgel Particles},
JOURNAL = {Polymers},
VOLUME = {3},
YEAR = {2011},
NUMBER = {4},
PAGES = {1575--1590},
URL = {https://www.mdpi.com/2073-4360/3/4/1575},
DOI = {10.3390/polym3041575}
}

@article{Ahmed2009minimum,
author = {Ahmed, Zeeshan and Gooding, Edward A. and Pimenov, Konstantin V. and Wang, Luling and Asher, Sanford A.},
title = {UV Resonance Raman Determination of Molecular Mechanism of Poly($N$-isopropylacrylamide) Volume Phase Transition},
journal = {J. Phys. Chem. B},
volume = {113},
number = {13},
pages = {4248-4256},
year = {2009},
doi = {10.1021/jp810685g}
}

@article{Yang2016ipnvsco,
    author = {Yang, Man and Zhao, Kongshuang},
    title = {Influence of the structure on the collapse of poly($N$-isopropylacrylamide)-based microgels: an insight by quantitative dielectric analysis},
    journal = {Soft Matter},
    volume = {12},
    number = {18},
    pages = {4093-4102},
    year = {2016},
    month = {05},
    issn = {1744-683X},
    doi = {10.1039/c5sm03155a}
}

@article{Stieger2004Rheo,
author = {Stieger, Markus and Pedersen, Jan Skov and Lindner, Peter and Richtering, Walter},
title = {Are Thermoresponsive Microgels Model Systems for Concentrated Colloidal Suspensions? A Rheology and Small-Angle Neutron Scattering Study},
journal = {Langmuir},
volume = {20},
number = {17},
pages = {7283-7292},
year = {2004},
doi = {10.1021/la049518x},
URL = {https://doi.org/10.1021/la049518x},

}

@article{Borue1988theory,
author = {Borue, V. Yu. and Erukhimovich, I. Ya.},
title = {A statistical theory of weakly charged polyelectrolytes: fluctuations, equation of state and microphase separation},
journal = {Macromolecules},
volume = {21},
number = {11},
pages = {3240-3249},
year = {1988},
doi = {10.1021/ma00189a019}
}

@article{Bhol2020catalyst,
doi = {10.1088/1361-648X/abbe79},
url = {https://doi.org/10.1088/1361-648X/abbe79},
year = {2020},
month = {dec},
publisher = {IOP Publishing},
volume = {33},
number = {8},
pages = {084002},
author = {Bhol, Prachi and Mohanty, Priti S},
title = {Smart microgel-metal hybrid particles of {PNIPAM-co-PAA@AgAu}: synthesis, characterizations and modulated catalytic activity},
journal = {J. Phys. Condens. Matter}
}

@article{Kureha2016separation,
author = {Kureha, Takuma and Shibamoto, Takahisa and Matsui, Shusuke and Sato, Takaaki and Suzuki, Daisuke},
title = {Investigation of Changes in the Microscopic Structure of Anionic Poly($N$-isopropylacrylamide-co-Acrylic acid) Microgels in the Presence of Cationic Organic Dyes toward Precisely Controlled Uptake/Release of Low-Molecular-Weight Chemical Compound},
journal = {Langmuir},
volume = {32},
number = {18},
pages = {4575-4585},
year = {2016},
doi = {10.1021/acs.langmuir.6b00760}
}

@article{Komarova2025emulsifier,
  author = {Komarova, Galina A. and Kozhunova, Elena Yu. and Gumerov, Rustam A. and Potemkin, Igor I. and Nasimova, Irina R.},
  title = {Effect of Polymer Network Architecture on Adsorption Kinetics at Liquid–Liquid Interfaces: A Comparison Between Poly({NIPAM-co-AA}) Copolymer Microgels and Interpenetrating Network Microgels},
  journal = {Gels},
  volume = {11},
  number = {1},
  pages = {58},
  year = {2025},
  doi = {10.3390/gels11010058},
  url = {https://www.mdpi.com/2310-2861/11/1/58},
  issn = {2310-2861},
  
}

@article{Buratti2020film,
  author = {Buratti, Elena and Sanzari, Ilaria and Dinelli, Franco and Prodromakis, Themistoklis and Bertoldo, Monica},
  title = {Formation and Stability of Smooth Thin Films with Soft Microgels Made of Poly($N$-Isopropylacrylamide) and Poly(Acrylic Acid)},
  journal = {Polymers},
  volume = {12},
  number = {11},
  pages = {2638},
  year = {2020},
  doi = {10.3390/polym12112638},
  url = {https://www.mdpi.com/2073-4360/12/11/2638},
  issn = {2073-4360},
}

@article{Burmistrova_Crosslinker_2011,
author = {Burmistrova, Anna and Richter, Marcel and {\"U}z{\"u}m, Cagri and von Klitzing, Regine},
year = {2011},
month = {04},
pages = {613-624},
title = {Effect of cross-linker density of P({NIPAM-co-AAc}) microgels at solid surfaces on the swelling/shrinking behaviour and the {Young’s} modulus},
volume = {289},
journal = {Colloid Polym. Sci.},
doi = {10.1007/s00396-011-2383-2}
}

@article{Witt2019,
author = {Witt, Marcus
U. and Hinrichs, Stephan and M{\"o}ller, Nadir and Backes, Sebastian and Fischer, Birgit and von Klitzing, Regine},
title = {Distribution of {CoFe2O4} Nanoparticles Inside {PNIPAM-Based} Microgels of Different Cross-linker Distributions},
journal = {J. Phys. Chem. B},
volume = {123},
number = {10},
pages = {2405-2413},
year = {2019},
doi = {10.1021/acs.jpcb.8b09236},
URL = {https://doi.org/10.1021/acs.jpcb.8b09236}, 

}

@article{Dongxia2006,
author = {Dongxia Huo and Yana Li and Qingwen Qian and Takaomi Kobayashi},
title = {Temperature–{pH} sensitivity of bovine serum albumin protein-microgels based on cross-linked poly($N$-isopropylacrylamide-co-acrylic acid)},
journal = {Colloids and Surfaces B: Biointerfaces},
volume = {50},
number = {1},
pages = {36-42},
year = {2006},
issn = {0927-7765},
doi = {https://doi.org/10.1016/j.colsurfb.2006.03.020},
url = {https://www.sciencedirect.com/science/article/pii/S092777650600110X},

}

@article{Reufer2009,
author  = {Reufer, M. and D{\'i}az-Leyva, P. and Lynch, I. and Scheffold, F.},
title   = {Temperature-sensitive {poly($N$-Isopropyl-Acrylamide)} microgel particles: A light scattering study},
journal = {Eur. Phys. J. E},
year    = {2009},
volume  = {28},
number  = {2},
pages   = {165--171},
url     = {https://doi.org/10.1140/epje/i2008-10387-2},
issn    = {1292-895X}
}

@manual{BAMM2013AcrylicAcid,
  author       = {{Basic Acrylic Monomer Manufacturers, Inc.}},
  title        = {Acrylic Acid: A Summary of Safety and Handling},
  edition      = {4},
  year         = {2013},
  organization = {Basic Acrylic Monomer Manufacturers, Inc.},
  address      = {USA},
  url          = {https://www.bamm.net/safe-handling-manuals},
}

@article{Rumyantsev2015segregation,
    author = {Rumyantsev, Artem M. and Rudov, Andrey A. and Potemkin, Igor I.},
    title = {Communication: Intraparticle segregation of structurally homogeneous polyelectrolyte microgels caused by long-range {Coulomb} repulsion},
    journal = {J. Chem. Phys.},
    volume = {142},
    number = {17},
    pages = {171105},
    year = {2015},
    month = {05},
    issn = {0021-9606},
    doi = {10.1063/1.4919951},
    url = {https://doi.org/10.1063/1.4919951}
}

@article{Schmidt2008,
    author = {Schmidt, Stephan and Hellweg, Thomas and von Klitzing, Regine},
    title = {Packing Density Control in {P(NIPAM-co-AAc)} Microgel Monolayers: Effect of Surface Charge, pH, and Preparation Technique},
    journal = {Langmuir},
    volume = {24},
    number = {21},
    pages = {12595-12602},
    year = {2008},
    month = {11},
    issn = {0743-7463},
    doi = {10.1021/la801770n},
    url = {https://doi.org/10.1021/la801770n}
}

\end{document}